\documentclass[onecolumn,12pt,showpacs,preprintnumbers,amsmath,amssymb,floatfix]{revtex4}

\usepackage{graphicx}
\usepackage{dcolumn}
\usepackage{bm}
\usepackage{amsfonts}

\DeclareMathAlphabet{\mathsfsl}{OT1}{cmr}{bx}{it}
\begin{document}
\title{Effect of surface roughness on rate-dependent slip in simple fluids}
\author{Nikolai~V.~Priezjev}

\homepage[]{http://www.egr.msu.edu/~priezjev}

\affiliation{Department of Mechanical Engineering, Michigan State
University, East Lansing, Michigan 48824}

\date{\today}
%
\begin{abstract}

Molecular dynamics simulations are used to investigate the influence
of molecular-scale surface roughness on the slip behavior in thin
liquid films. The slip length increases almost linearly with the
shear rate for atomically smooth rigid walls and incommensurate
structures of the liquid/solid interface. The thermal fluctuations
of the wall atoms lead to an effective surface roughness, which
makes the slip length weakly dependent on the shear rate. With
increasing the elastic stiffness of the wall, the surface roughness
smoothes out and the strong rate dependence is restored again. Both
periodically and randomly corrugated rigid surfaces reduce the slip
length and its shear rate dependence.

\end{abstract}



\maketitle

\section{Introduction}


The description of the fluid flow in confined geometry requires
specification of the boundary condition for the fluid velocity at
the solid wall. Usually the fluid is assumed to be immobile at the
boundary. Although this assumption is successful in describing fluid
flow on macroscopic length scales, it needs a revision for the
microscopic scales due to possible slip of the fluid relatively to
the wall~\cite{KarniBeskok}. The existence of liquid slip at the
solid surfaces was established in many experiments on the pressure
driven flow in narrow
capillaries~\cite{Schnell56,Churaev84,Breuer03} and drainage of thin
liquid films in the surface force
apparatus~\cite{MackayVino,Charlaix01,Granick01}. The most popular
Navier model relates the fluid slip velocity to the interfacial
shear rate by introducing the slip length, which is assumed to be
rate-independent. The slip length is defined as a distance from the
boundary where the linearly extrapolated fluid velocity profile
vanishes. Typical values of the slip length inferred from the
experiments on fluids confined between smooth hydrophobic surfaces
is of the order of ten
nanometers~\cite{Granick01,MackayVino,Charlaix05,BocquetPRL06,Vinograd06}.
Despite the large amount of experimental data on the slip
length~\cite{BonaccursoRev05}, the underlying molecular mechanisms
leading to slip are still poorly understood because it is very
difficult to resolve the fluid velocity profile in the region near
the liquid/solid interface at these length scales.

Over the last twenty years, molecular dynamics (MD) simulations were
extensively used to investigate the correlation between the
structure of simple fluids in contact with atomically smooth
surfaces and slip boundary
conditions~\cite{KB89,Thompson90,Barrat94,Barrat99,Travis00,Cieplak01,Quirke01,Khare06}.
The advantage of the MD method is that the fluid velocity profile
can be resolved at the molecular level and no assumptions about the
slip velocity at the interface are required. The main factors
affecting slip for atomically smooth surfaces are the wall-fluid
interaction, the degree of commensurability of liquid and solid
structures at the interface, and diffusion of fluid molecules near
the wall. The slip length was found to correlate inversely with the
wall-fluid interaction energy and the amount of structure induced in
the first fluid layer by the periodic surface
potential~\cite{Thompson90}. For weak wall-fluid interactions and
smooth surfaces, the slip length is proportional to the collective
relaxation coefficient of the fluid molecules near the
wall~\cite{Barrat99fd}. The thermal fluctuations of the wall atoms
under the strong harmonic potential reduce the degree of the
in-plane fluid ordering and result in larger values of the slip
length~\cite{Thompson90}. On the other hand, an excessive
penetration of the wall atoms into the fluid phase reduces the slip
velocity for soft thermal walls~\cite{Tanner99}. Nevertheless, the
effect of thermal surface roughness on the slip length in the
\textit{shear-rate-dependent} regime was not systematically explored
even for atomically smooth walls.

In the original MD study by Thompson and Troian~\cite{Nature97} on
boundary driven shear flow of simple fluids past atomically smooth
rigid walls, the slip length was found to increase nonlinearly with
the shear rate for weak wall-fluid interactions. A similar dynamic
behavior of the slip length has been reported in thin polymer
films~\cite{Priezjev04}. The rate-dependent slip was also observed
for the planar Poiseuille flow of simple fluids confined between
hydrophobic surfaces with variable size of the wall
atoms~\cite{Fang05,Yang06}. The variation of the slip length (from
negative to positive values) with increasing shear rate for
hydrophilic surfaces~\cite{Fang05} can be well described by the
power law function proposed in Ref.~\cite{Nature97}. In the recent
paper~\cite{Priezjev07}, we have reported a gradual transition in
the shear rate dependence of the slip length, from linear to highly
nonlinear function with pronounced upward curvature, by decreasing
the strength of the wall-fluid interaction. Remarkably, in a wide
range of shear rates and surface energies, the slip length is well
fitted by a power law function of a single variable, which is a
combination of the structure factor, contact density, and
temperature of the first fluid layer. One of the goals of the
present study is to investigate how the rate-dependent slip is
affected by the presence of the molecular-scale surface roughness.

Molecular scale simulations of simple~\cite{Attard04,Priezjev06} and
polymeric~\cite{Gao2000,Jabb00} fluids (as well as recent
experiments~\cite{Granick02,Archer03,Leger06}) have shown that the
slip length is reduced in the presence of the surface roughness. The
effect is enhanced for smaller wavelengths and larger amplitudes of
the surface corrugation~\cite{Jabb00,Priezjev06}. At low shear
rates, the reduction of the effective slip length is caused by the
local curvature of the fluid flow above macroscopic surface
corrugations~\cite{Richardson73,Panzer90,Priezjev06} or by more
efficient trapping of the fluid molecules by atomic-scale surface
inhomogeneities~\cite{Thompson90,Barrat94,Attard04,Priezjev05,Priezjev06}.
The analysis of more complex systems with combined effects of
surface roughness and rate dependency poses certain difficulties in
the interpretation of the experimental results because the exact
dependence of the local slip length on shear rate is often not
known.

In this paper, we explore the influence of molecular-scale surface
roughness on the slip behavior in a flow of simple fluids driven by
a constant force. We will show that the functional form of the
rate-dependent slip length is considerably modified by the presence
of the thermal, random and periodic surface roughness. The growth of
the slip length with increasing shear rate, which is observed for
atomically smooth rigid walls, is strongly reduced by periodic and
random surface roughness. Soft thermal walls produce very weak rate
dependence of the slip length, while the linear behavior is restored
for stiffer walls.

The paper is organized as follows. The details of molecular dynamics
simulations are described in the next section. Results for the shear
rate dependence of the slip length on the thermal and random surface
roughness are presented in Section~\ref{sec:Results_thermo}. The
effect of periodic wall roughness on the rate-dependent slip is
discussed in Section~\ref{sec:Results_corrugated}. The summary is
given in the last section.

\section{MD Simulation model}
\label{sec:Model}

The simulation setup consists of $N\,{=}\,3456$ fluid molecules
confined between two stationary atomistic walls parallel to the $xy$
plane. The molecules interact through the truncated Lennard-Jones
(LJ) potential
\begin{equation}
V_{LJ}(r)\!=4\,\varepsilon\,\Big[\Big(\frac{\sigma}{r}\Big)^{12}\!-\Big(\frac{\sigma}{r}\Big)^{6}\,\Big],
\end{equation}
where $\varepsilon$ and $\sigma$ represent the energy and length
scales of the fluid phase with density
$\rho\,{=}\,0.81\,\sigma^{-3}$. The interaction between wall atoms
and fluid molecules is also modeled by the LJ potential with the
energy $\varepsilon_{\rm wf}$ and length scale $\sigma_{\rm wf}$
measured in units of $\varepsilon$ and $\sigma$. In all our
simulations, wall atoms do not interact with each other and
$\sigma_{\rm wf}\,{=}\,\sigma$. The cutoff distance is set to
$r_c\,{=}\,2.5\,\sigma$ for fluid-fluid and wall-fluid interactions.

\begin{table}[b]
\caption{Root mean-square displacement, $\langle\delta
u^2\rangle\,{=}\,3\,k_BT/\kappa$, divided by the nearest-neighbor
distance $d\,{=}\,0.8\,\sigma$ and the typical oscillation time of
the wall atoms tethered about their equilibrium lattice positions as
a function of the spring stiffness for $m_w\,{=}\,4\,m$.}
 \vspace*{3mm}
 \begin{ruledtabular}
 \begin{tabular}{r r r r r r}
     Spring stiffness \\
     $\kappa\,[\varepsilon/\sigma^2]~~~~~$ & $400\,$ & $600\,$ & $800\,$ & $1200$ & $1600$
     \\ [3pt] \hline \\[-5pt]
     $\sqrt{\langle\delta u^2\rangle}/d~~~~$ & $0.11$ & $0.09$ & $0.08$ & $0.07$ & $0.06$
     \\ [3pt]
     $2\pi\sqrt{m_w/\kappa}\,\,[\tau]$  &      $0.63$ &  $0.51$ & $0.44$ & $0.36$ & $0.31$
     \\ [2pt]
 \end{tabular}
 \end{ruledtabular}
 \label{tabela}
\end{table}

The steady-state flow was induced by a constant force
$\text{f}_{\text{x}}$ in the $\hat{x}$ direction, which acted on
each fluid molecule. The heat exchange with external reservoir was
controlled by a Langevin thermostat with a random, uncorrelated
force and a friction term, which is proportional to the velocity of
the fluid molecule. The value of the friction coefficient
$\Gamma\,{=}\,1.0\,\tau^{-1}$ is small enough not to affect
significantly the self-diffusion coefficient of the fluid
molecules~\cite{Grest86,GrestJCP04}. The Langevin thermostat was
applied only along the $\hat{y}$ axis to avoid a bias in the flow
direction~\cite{Thompson90}. The equations of motion for a fluid
monomer of mass $m$ are given by
\begin{eqnarray}
\label{Langevin_x}
m\ddot{x}_i & = & -\sum_{i \neq j} \frac{\partial V_{ij}}{\partial x_i} + \text{f}_{\text{x}}\,, \\
\label{Langevin_y}
m\ddot{y}_i + m\Gamma\dot{y}_i & = & -\sum_{i \neq j} \frac{\partial V_{ij}}{\partial y_i} + f_i\,, \\
\label{Langevin_z}
m\ddot{z}_i & = & -\sum_{i \neq j} \frac{\partial V_{ij}}{\partial z_i}\,, %
\end{eqnarray}
where $f_i$ is a randomly distributed force with $\langle
f_i(t)\rangle\,{=}\,0$ and variance $\langle
f_i(0)f_j(t)\rangle\,{=}\,\,2mk_BT\Gamma\delta(t)\delta_{ij}$
obtained from the fluctuation-dissipation theorem. The temperature
of the thermostat is fixed to $T\,{=}\,1.1\,\varepsilon/k_B$, where
$k_B$ is the Boltzmann constant. The equations of motion of the
fluid molecules and wall atoms are integrated using the fifth-order
gear-predictor method~\cite{Allen87} with a time step $\triangle
t\,{=}\,0.002\,\tau$, where
$\tau\,{=}\,\sqrt{m\sigma^2/\varepsilon}$ is the LJ time.

The wall atoms were allowed to oscillate about their equilibrium
lattice sites under the harmonic potential
$V_{sp}\,{=}\,\frac{1}{2}\,\kappa\,r^2$. The spring stiffness
$\kappa$ was chosen so that the ratio of the root mean-square
displacement of the wall atoms and their nearest-neighbor distance
was less than the Lindemann criterion for melting,
$\sqrt{\langle\delta u^2\rangle}/d\lesssim 0.15$, e.g. see
Ref.~\cite{Barrat03}. At the same time, the parameter $\kappa$
should be small enough so that the dynamics of the wall atoms can be
accurately resolved with the MD integration time step. The mass of
the wall atoms $m_w$ was chosen to be four times that of the fluid
molecule to make their oscillation times comparable. The Langevin
thermostat was also applied to all three components of the wall
atoms equations of motion, e.g. for the $\hat{x}$ component
\begin{eqnarray}
\label{Langevin_wall_x} m_w\,\ddot{x}_i + m_w\,\Gamma\dot{x}_i & = &
-\sum_{i \neq j} \frac{\partial V_{ij}}{\partial x_i}
- \frac{\partial V_{sp}}{\partial x_i} + f_i\,, 
\end{eqnarray}
where the sum is taken over the fluid molecules within the cutoff
radius $r_c\!\,\,{=}\,\,2.5\,\sigma$. Mean displacement of the wall
atoms and their typical oscillation time are summarized in
Table~\ref{tabela} for the values of the spring constant considered
in this study.

Each wall is constructed out of $648$ atoms distributed between two
(111) planes of the face-centered cubic (fcc) lattice with density
$\rho_w\,{=}\,2.73\,\sigma^{-3}$. The walls are separated by a
distance $24.58\,\sigma$ along the $\hat{z}$ axis. The lateral
dimensions of the computational domain in the $xy$ plane are
measured as $L_x\,{=}\,25.03\,\sigma$ and $L_y\,{=}\,7.22\,\sigma$.
Periodic boundary conditions are imposed along the $\hat{x}$ and
$\hat{y}$ directions. After an equilibration period of at least
$2\times10^4\tau$, the fluid velocity and density profiles were
computed for a time interval up to $2\times10^5\tau$ within bins of
thickness $\Delta z\,{=}\,0.2\,\sigma$ and $\Delta
z\,{=}\,0.01\,\sigma$, respectively~\cite{Priezjev07}. The shear
viscosity, $\mu\,{=}\,(2.2\pm0.2)\,\,\varepsilon\tau\sigma^{-3}$,
remained independent of the external
force~\cite{Priezjev04,Priezjev07}.

\begin{figure}[t]
\includegraphics[width=10.4cm,height=7.6cm,angle=0]{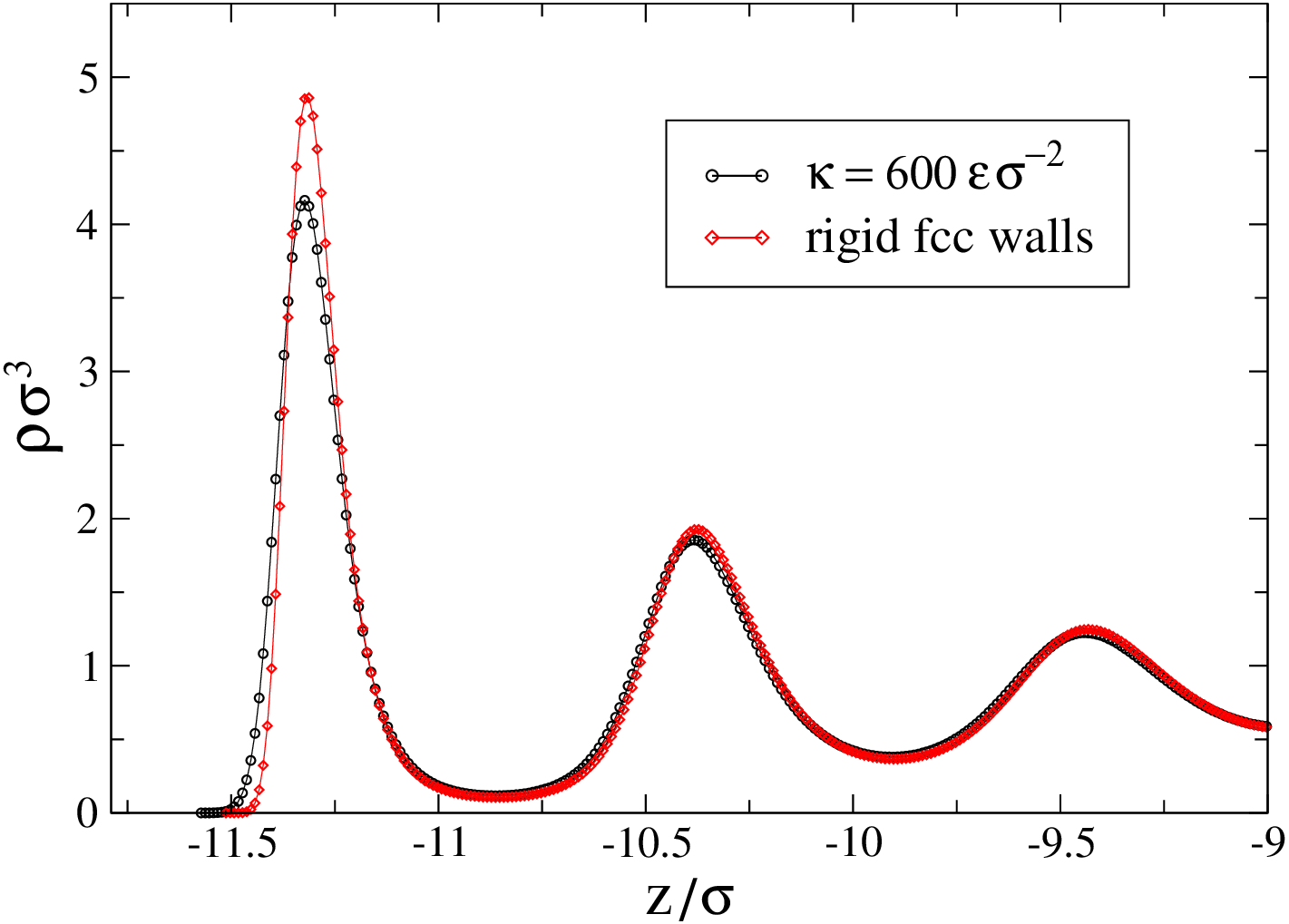}
\caption{(Color online) Averaged fluid density profiles near the
thermal $\kappa\,{=}\,600\,\varepsilon/\sigma^2$ ($\circ$) and rigid
($\diamond$) walls for $\varepsilon_{\rm wf}/\varepsilon\,{=}\,0.9$
and $\text{f}_{\text{x}}\,{=}\,0.001\,\varepsilon/\sigma$. Left
vertical axis coincides with the location of the liquid/solid
interface at $z=-11.79\,\sigma$.} \label{mol_dens}
\end{figure}

\section{Results for the thermal walls}
\label{sec:Results_thermo}

\subsection{Fluid density and velocity profiles}

Examples of the averaged fluid density profiles near the thermal and
rigid walls are presented in Fig.\,\ref {mol_dens} for a small value
of the external force
$\text{f}_{\text{x}}\,{=}\,0.001\,\varepsilon/\sigma$. The first
peak in the density profile is slightly broader for the thermal
walls because the fluid molecules can move closer to the fcc lattice
plane due to finite spring stiffness of the wall atoms. The maximum
value of the first peak defines a contact density, which is larger
for the rigid walls because of the higher in-plane fluid ordering.
In both cases, the fluid density oscillations gradually decay to a
uniform bulk profile within $5\!-\!6 \,\sigma$ away from the wall
(not shown). The contact density decreases slightly for larger
values of the applied force. A correlation between the fluid
structure near the wall and the slip length will be examined in the
next section.

\begin{figure}[t]
\includegraphics[width=10.4cm,height=7.6cm,angle=0]{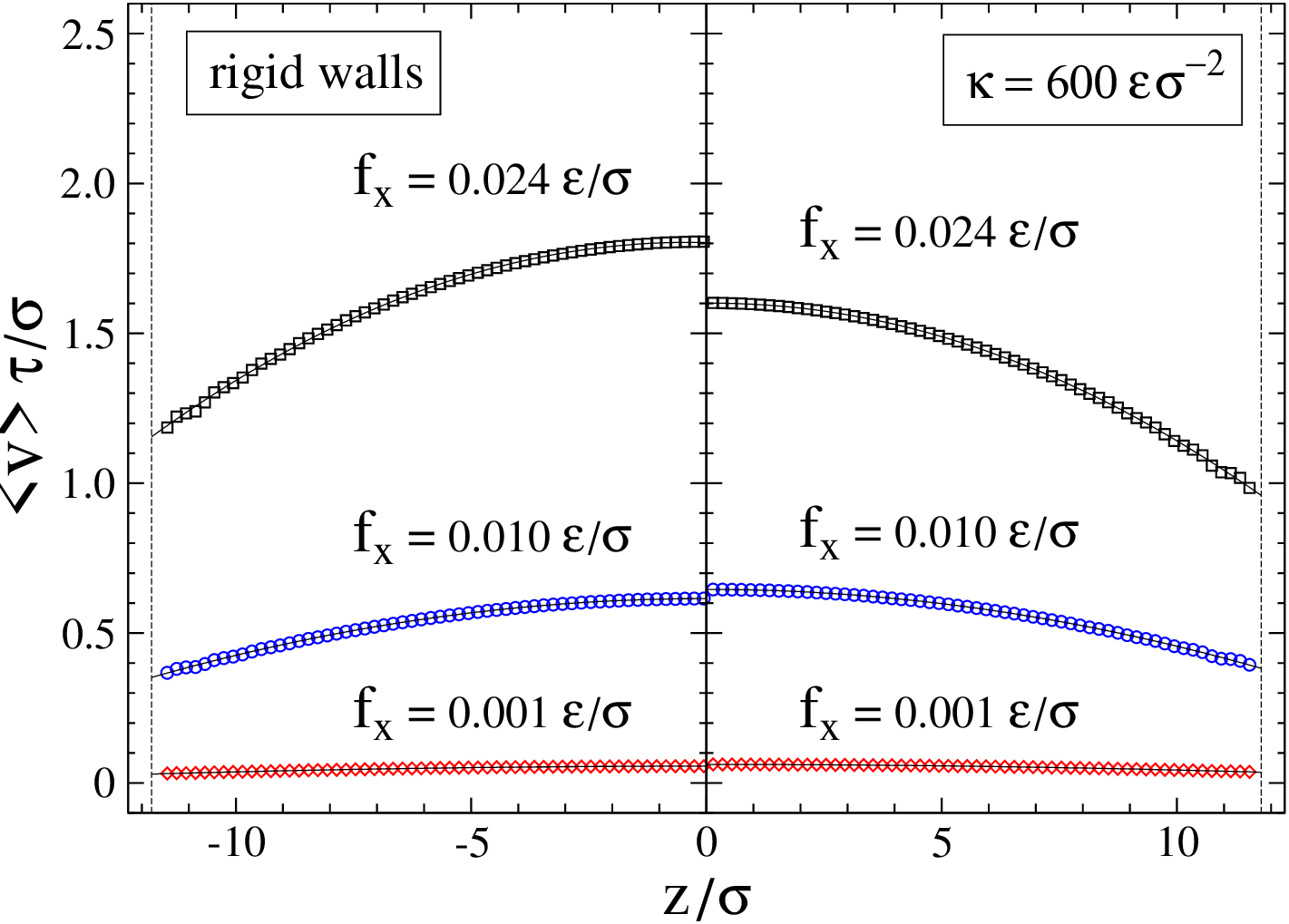}
\caption{(Color online) Averaged velocity profiles, $\langle
\text{v} \rangle\, \tau/\sigma$, for the indicated values of the
applied force per fluid molecule for the rigid walls (left) and the
thermal walls with $\kappa\,{=}\,600\,\varepsilon/\sigma^2$ (right).
The wall-fluid interaction energy is fixed to $\varepsilon_{\rm
wf}/\varepsilon\,{=}\,0.9$. The solid lines represent a parabolic
fit to the data. The dashed lines indicate the location of
liquid/solid interfaces at $z=\pm\,11.79\,\sigma$. Vertical axes
denote the position of the fcc lattice planes at $z=\pm
12.29\,\sigma$.} \label{parab_velo}
\end{figure}

The solution of the Navier-Stokes equation for the force driven flow
with slip boundary conditions at the confining parallel walls,
$\textrm{v}(\pm h)\,{=}\,V_s$, is given by~\cite{KarniBeskok}
\begin{equation}
\text{v}(z)=\frac{\rho\,\text{f}_{\text{x}}}{2\mu}\,(h^2-z^2)+V_s\,,
\label{velo_hydro}
\end{equation}
where $2h$ is the distance between the walls and $\mu$ is the shear
rate independent viscosity. The slip length is defined as an
extrapolated distance relative to the position of the liquid/solid
interface where the tangential component of the fluid velocity
vanishes
\begin{equation}
\Big|\frac{\partial \text{v}}{\partial z}\,(\pm
h)\Big|=\frac{V_s}{L_s}\,.
\end{equation}

Figure\,\ref{parab_velo} shows representative velocity profiles in
steady-state flow for three different values of the external force
$\text{f}_{\text{x}}$ and fixed wall-fluid interaction energy
$\varepsilon_{\rm wf}/\varepsilon\,{=}\,0.9$. The data for thermal
and rigid walls are presented only in half of the channel because of
the symmetry with respect to the mid-plane of the fluid phase, i.e.
$\text{v}(z)\!=\text{v}(-z)$. Fluid velocity profiles are well
fitted by a parabola with a shift by the value of the slip velocity,
as expected from the continuum predictions [see
Eq.\,(\ref{velo_hydro})]. The simulation results presented in
Fig.\,\ref{parab_velo} show that slip velocity $V_s$ increases with
the applied force. The degree of slip depends on the wall stiffness
and the interfacial shear rate. The fluid slip velocity is larger
for the thermal walls and small forces
$\text{f}_{\text{x}}\,{\leqslant}\,0.012\,\varepsilon/\sigma$. By
contrast, rigid walls produce more slippage for the large value of
the external force
$\text{f}_{\text{x}}\,{=}\,0.024\,\varepsilon/\sigma$. In both
cases, the slip velocity is greater than the difference between the
fluid velocities at the center of the channel and near the walls.
The upper bound for the Reynolds number is
$Re\approx10$~\cite{Priezjev07}, ensuring laminar flow conditions
throughout.

\subsection{Effect of thermal wall roughness on slip length}

Slip boundary conditions for a fluid flow past atomically smooth
rigid walls are determined by the molecular-scale surface roughness
due to the wall atoms fixed at their equilibrium lattice sites. The
thermal fluctuations of the wall atoms, being unavoidable in real
surfaces, modify the effective coupling between liquid and solid
phases. Depending on the stiffness of the surface, thermal walls can
either reduce slip (due to the deep penetration of the wall atoms
into the fluid phase) or increase slip because of the reduction of
the surface induced structure in the adjacent fluid layer. In this
section, the shear rate dependence of the slip length is
investigated in a wide range of values of the parameter $\kappa$
that satisfy the Lindemann criterion for melting (see
Table~\ref{tabela}).

In the previous MD study on shear flow near solids by Thompson and
Robbins~\cite{Thompson90}, it was shown that the thermal surface
roughness reduces the degree of ordering in the adjacent fluid
layer. The thermal fluctuations of the wall atoms produced slip
lengths of about $0.5\,\sigma$ larger than their values for the
rigid walls. In a wide range of wall-fluid interaction energies
$0.2\,{\leqslant}\,\varepsilon_{\rm
wf}/\varepsilon\,{\leqslant}\,25$, the slip length was found to be
rate-independent and less than $3.5\,\sigma$. In our simulations,
the surface potential is less corrugated because of the higher wall
density; and, therefore, the effect of thermal surface roughness on
the slip length is greater.

\begin{figure}[t]
\includegraphics[width=10.4cm,height=7.6cm,angle=0]{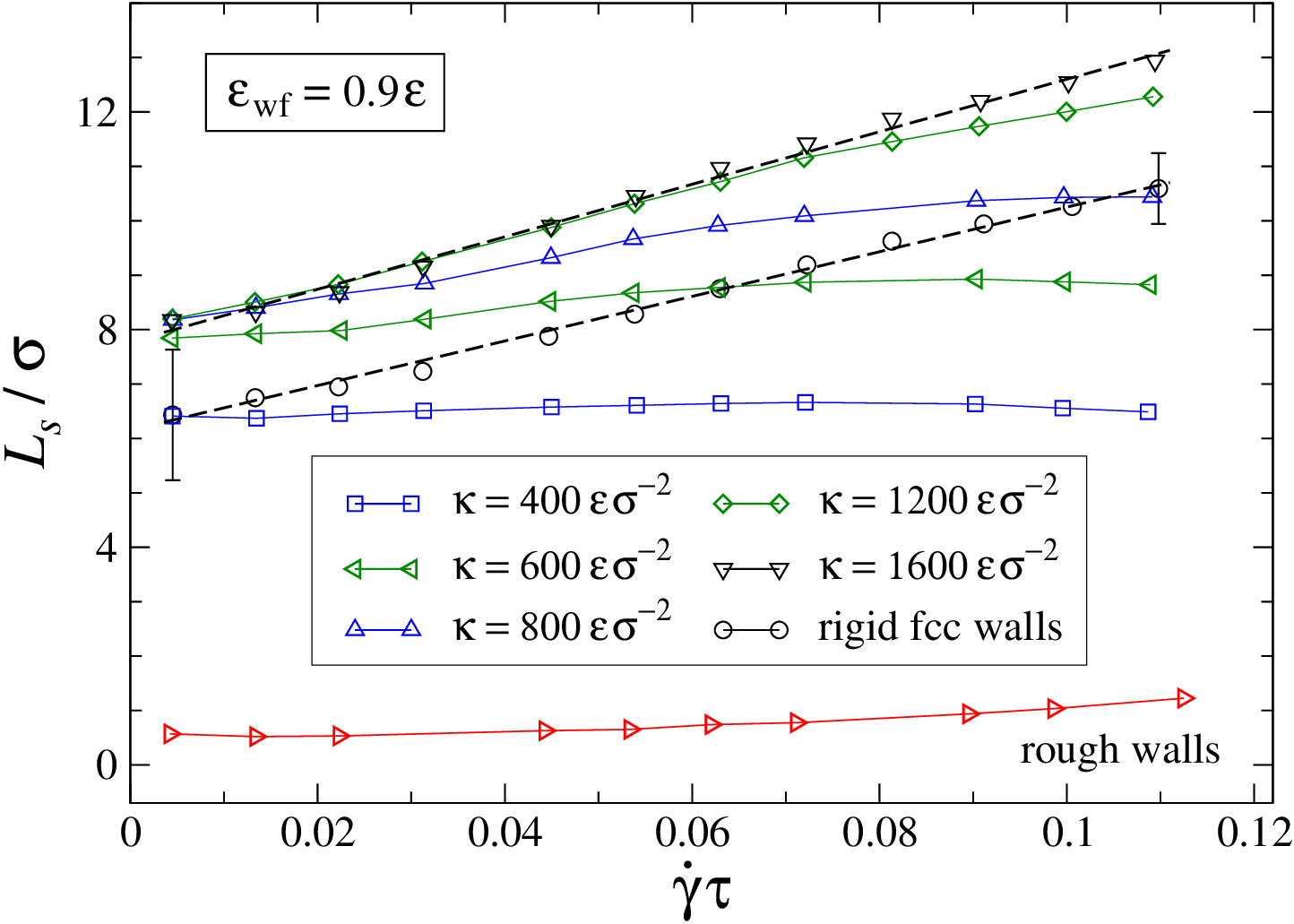}
\caption{(Color online) Slip length, $L_s/\sigma$, as a function of
the shear rate at the interface for $\varepsilon_{\rm
wf}/\varepsilon\,{=}\,0.9$. The values of the spring constant
$\kappa$ for the thermal walls are listed in the inset. The dashed
lines represent the best fit to the data for the thermal walls with
$\kappa\,{=}\,1600\,\varepsilon/\sigma^2$ ($\triangledown$) and the
rigid fcc walls ($\circ$). The slip length for rough rigid walls
with $\langle\delta u^2\rangle^{1/2}\!\simeq0.07\,\sigma$
($\triangleright$). The solid curves are a guide for the eye.}
\label{thermo}
\end{figure}

The variation of the slip length with increasing shear rate for
different values of the spring stiffness $\kappa$ and
$\varepsilon_{\rm wf}/\varepsilon\,{=}\,0.9$ is presented in
Fig.\,\ref{thermo}. The data for smooth rigid walls, fitted by a
straight line, are also shown in Fig.\,\ref{thermo} for comparison
with the results for the thermal walls. In the range of accessible
shear rates, the slip length is larger for stiffer thermal walls.
The surface becomes effectively smoother because the average
displacement of the wall atoms with respect to their equilibrium
sites is reduced at larger values of $\kappa$. A similar reduction
in slip velocity for the soft thermal wall atoms was reported in
recent MD simulations of thin films of hexadecane~\cite{Tanner99}.
For soft walls with $\kappa\,{=}\,400\,\varepsilon/\sigma^2$, the
wall atoms penetrate deeper into the fluid phase, which makes the
slip length smaller and weakly dependent on the shear rate. For
$\kappa\,{=}\,600\,\varepsilon/\sigma^2$, the slip length increases
slightly at low shear rates and then saturates at
$\dot{\gamma}\tau\!\gtrsim0.063$, where it becomes smaller than
$L_s$ for the rigid walls. This behavior is consistent with the
results for the fluid velocity profiles presented in
Fig.\,\ref{parab_velo} for the thermal and rigid walls.

In the case of the largest spring constant
$\kappa\,{=}\,1600\,\varepsilon/\sigma^2$, the slip length increases
monotonically with the shear rate and its dependence can also be
fitted well by a straight line. The slope of the fitted line is
slightly larger than one for the rigid walls (see
Fig.\,\ref{thermo}). For a finite spring stiffness, a small downward
curvature appears at $\dot{\gamma}\tau\!\gtrsim0.05$ because of the
higher temperature and, as a consequence, larger mean displacement
of the wall atoms. The maximum increase in temperature of the wall
atoms and the adjacent fluid layer is about $10\%$ at the highest
shear rates reported in Fig.\,\ref{thermo}. The fluid temperature at
the center of the channel increases up to $1.106\,\varepsilon/k_B$
at the highest $\dot{\gamma}$. The upper bound for the shear rate is
determined by the maximum shear stress the liquid/solid interface
can support~\cite{Priezjev07}. The simulation results for the
thermal walls presented in Fig.\,\ref{thermo} demonstrate that the
spring stiffness in the model of harmonic oscillators can be an
important factor in determining the degree of slip in the
rate-dependent regime.

\begin{figure}[t]
\includegraphics[width=10.4cm,height=7.6cm,angle=0]{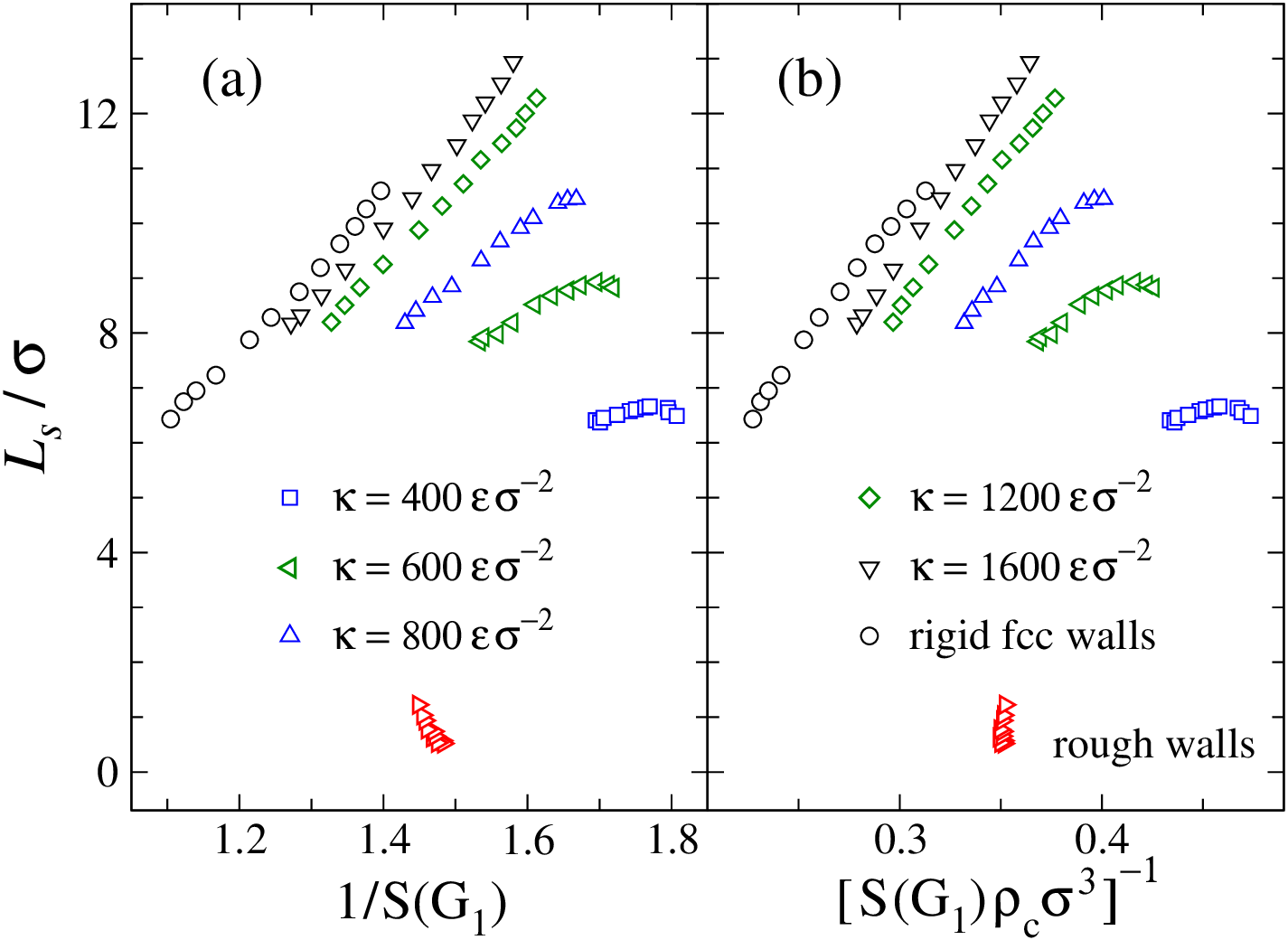}
\caption{(Color online) (a) Behavior of the slip length $L_s/\sigma$
as a function of the inverse value of the in-plane structure factor,
$1/S(\mathbf{G}_1)$, evaluated at the first reciprocal lattice
vector $\mathbf{G}_1\,{=}\,(9.04\,\sigma^{-1},0)$. The wall-fluid
interaction energy is $\varepsilon_{\rm wf}/\varepsilon\,{=}\,0.9$.
(b) The same data for the indicated values of the spring constant
are replotted versus $[S(\mathbf{G}_1)\,\rho_c\,\sigma^3]^{-1}$.}
\label{ls_S9_ro}
\end{figure}

It is interesting to note that the surface roughness due to immobile
wall atoms with random displacement of only a fraction of a
molecular diameter significantly reduces slip length and leads to a
slight upward curvature in the rate dependence (see
Fig.\,\ref{thermo}). The rough surfaces were constructed by fixing
the instantaneous positions of initially equilibrated wall atoms
with the spring stiffness $\kappa\,{=}\,\,600\,\varepsilon/\sigma^2$
in the absence of the flow. The parabolic velocity profiles for
different values of the applied force were computed for the same
realization of disorder. These random perturbations of the surface
potential lead to the difference in the slip length of about
$7\,\sigma$ in comparison with its values for the thermal walls with
the spring stiffness $\kappa\,{=}\,\,600\,\varepsilon/\sigma^2$.

The degree of slip at the liquid/solid interface correlates well
with the amount of the surface induced order in the adjacent fluid
layer~\cite{Thompson90,Barrat99fd,Priezjev04,Priezjev05}. The
in-plane fluid structure factor is defined as
$S(\mathbf{k})\,{=}\,1/N_{\ell}\,\,|\sum_j
e^{i\,\mathbf{k}\cdot\mathbf{r}_j}|^2$, where
$\mathbf{r}_j\,{=}\,(x_j,y_j)$ is the two-dimensional position
vector of the $j$-th molecule and the sum is taken over $N_{\ell}$
molecules within the first layer. The effect of periodic surface
potential on the structure of the adjacent fluid becomes more
pronounced at the reciprocal lattice vectors. In the previous MD
study~\cite{Priezjev07} for similar parameters of the wall and fluid
phases, it was shown that the slip length scales as
$L_s\!\sim\!(\,T_1/S(\mathbf{G}_1)\,\rho_c)^{\alpha}$, where
$\mathbf{G}_1$ is the the first reciprocal lattice vector in the
flow direction, $T_1$ is temperature of the first fluid layer and
$\alpha\,{=}\,1.44\pm0.10$. This scaling relation was found to hold
in a wide range of shear rates and wall-fluid interactions for
atomically smooth rigid walls and incommensurate structures of the
liquid/solid interface~\cite{Priezjev07}.

The correlation between the inverse value of the fluid structure
factor evaluated at the first reciprocal lattice vector
$\mathbf{G}_1\,{=}\,(9.04\,\sigma^{-1},0)$ and the slip length is
presented in Fig.\,\ref{ls_S9_ro}\,(a). Except for the rough rigid
walls, the surface induced structure in the first fluid layer
$S(\mathbf{G}_1)$ is reduced at higher shear rates and smaller
values of $\kappa$. The slip length increases approximately linearly
with $1/S(\mathbf{G}_1)$ for the rigid and stiff walls with
$\kappa\geqslant1200\,\varepsilon/\sigma^2$. A gradual transition to
a weak dependence of the slip length on $S(\mathbf{G}_1)$ is
observed upon reducing the wall stiffness. In
Figure~\ref{ls_S9_ro}\,(b) the same data for the slip length are
replotted as a function of the inverse product
$[S(\mathbf{G}_1)\,\rho_c\,\sigma^3]^{-1}$, where $\rho_c$ is a
contact density of the first fluid layer. Although the contact
density decreases slightly with increasing the slip velocity, the
functional form of the slip length is similar in both cases [see
Figs.\,\ref{ls_S9_ro}\,(a)\,--\,\ref{ls_S9_ro}\,(b)]. The results
shown in Fig.\,\ref{ls_S9_ro} indicate that the dependence of the
slip length on the fluid structure of the first layer is less
pronounced for $\kappa\leqslant800\,\varepsilon/\sigma^2$ due the
thermal surface roughness in the rate-dependent regime. Whether the
scaling relation for the slip length~\cite{Priezjev07} holds in the
presence of the thermal roughness for different wall-fluid
interaction energies will be the subject of the future research.

\section{Results for periodically corrugated walls}
\label{sec:Results_corrugated}

Next, the results for the rate dependence of the slip length are
compared for atomically smooth rigid walls and periodically
roughened surfaces. The periodic surface roughness of the upper and
the lower walls was modeled by introducing a vertical offset to the
positions of the wall atoms $\Delta z(x)\,{=}\,a\sin(2\pi
x/\lambda)$ with the wavelength $\lambda\,{=}\,4.17\,\sigma$. In
this part of the study, the wall atoms are rigidly fixed with
respect to their equilibrium sites. To properly compare the results
for atomically smooth and roughened surfaces, both the local shear
rate and the slip length were estimated from a parabolic fit of the
velocity profiles at the same location of the interface, i.e.
$z=\pm\,11.79\,\sigma$. A weaker wall-fluid interaction,
$\varepsilon_{\rm wf}/\varepsilon\,{=}\,0.5$, was chosen to obtain
larger values of $L_s$ in the absence of the imposed corrugation,
since it is expected that the surface roughness strongly reduces the
slip length~\cite{Priezjev06}.

\begin{figure}[t]
\includegraphics[width=10.4cm,height=7.6cm,angle=0]{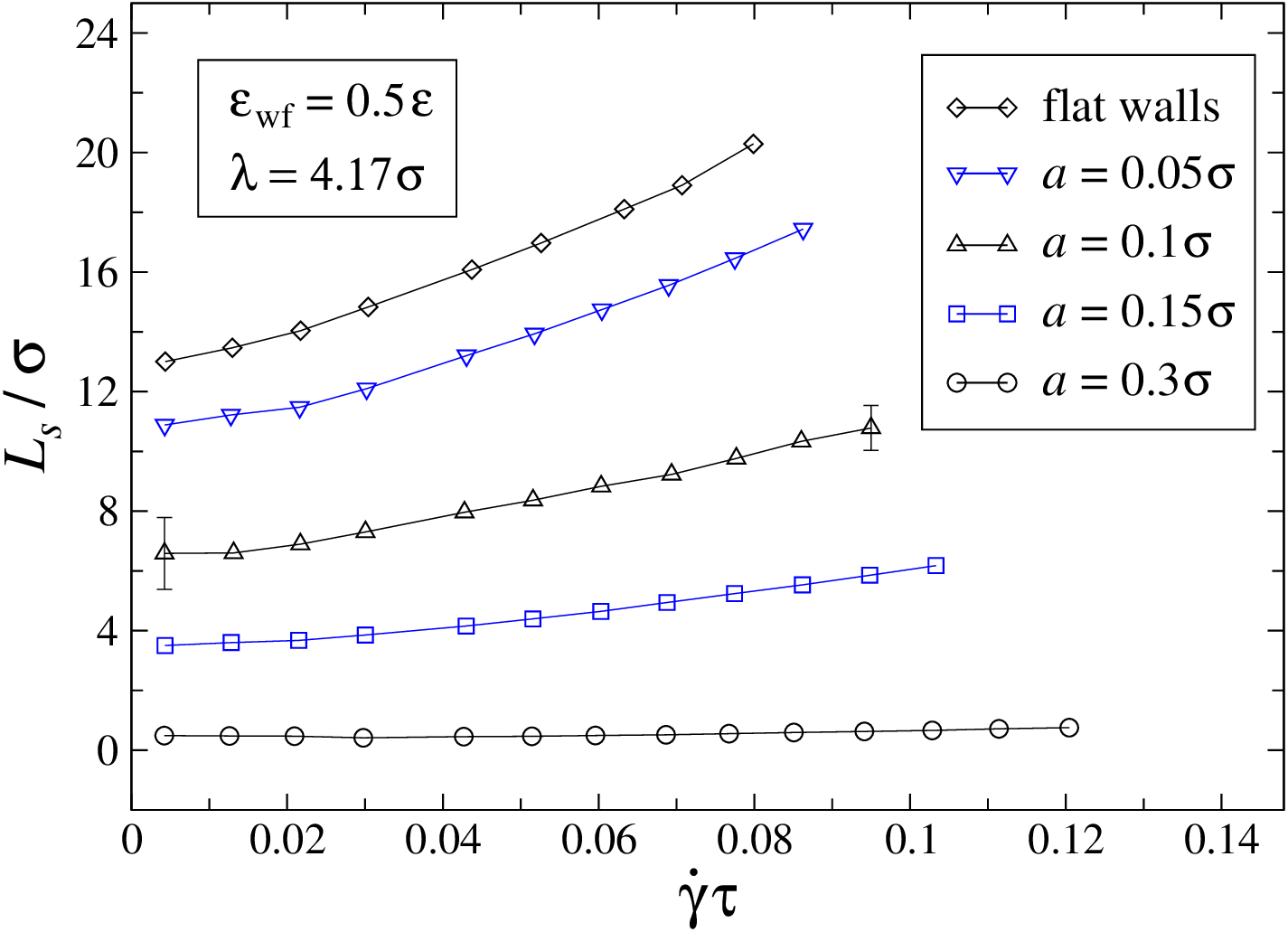}
\caption{(Color online) Variation of the slip length $L_s/\sigma$ as
a function of the local shear rate for Poiseuille flows and
$\varepsilon_{\rm wf}/\varepsilon\,{=}\,0.5$. The wavelength and
amplitudes of the wall modulation are tabulated in the insets. The
local shear rate and the slip length are extracted from a parabolic
fit of the velocity profiles at $z=\pm\,11.79\,\sigma$.}
\label{rough}
\end{figure}

The dynamic response of the slip length with increasing shear rate
is presented in Fig.\,\ref{rough} for atomically smooth and
periodically corrugated walls. The data for the flat walls
($a\,{=}\,0$) are the same as in Ref.~\cite{Priezjev07}. At low
shear rates, the slip length $L_s^{\circ}$ (defined by the leftmost
point on each curve in Fig.\,\ref{rough}) decreases monotonically
with increasing the amplitude $a$. The no-slip boundary condition is
achieved for $a\gtrsim0.3\,\sigma$. This behavior for a similar
geometry and interaction parameters was examined in detail in the
previous paper~\cite{Priezjev06}. A direct comparison between
continuum analysis and MD simulations showed that there is an
excellent agreement between the velocity profiles and the slip
length for the large wavelengths $\lambda\gtrsim20\,\sigma$ and
small values of $a/\lambda\lesssim0.05$. The continuum results
overestimate the slip length when $\lambda$ approaches molecular
dimensions~\cite{Priezjev06}.

At higher shear rates, the slope of the rate-dependent slip length
is gradually reduced with increasing the amplitude of the surface
corrugation (see Fig.\,\ref{rough}). For the largest amplitude
$a\,{=}\,0.3\,\sigma$, the slip length weakly depends on shear rate
and its magnitude becomes smaller than the molecular diameter. As
apparent from the set of curves shown in Fig.\,\ref{rough}, the same
value of the slip length can be obtained by increasing
simultaneously the amplitude of the surface corrugation and the
shear rate. Analogous behavior of the slip length was observed
experimentally for flows of Newtonian liquids past surfaces with
variable nano-roughness~\cite{Granick02}. We note, however, that the
MD simulations of simple fluids reported in this study do not show
any threshold in the rate dependence of the slip length for the
amplitudes of the surface corrugation $a\,{\leqslant}\,0.3\,\sigma$.

\section{Summary}
\label{sec:Conclusions}

In this paper the effect of molecular-scale surface roughness on the
slip length in a flow of simple fluids was studied by molecular
dynamics simulations. The parabolic fit of the steady-state velocity
profiles induced by a constant force was used to define the values
of interfacial shear rate and slip length. For atomically smooth
rigid surfaces and weak wall-fluid interactions, the slip length
increases approximately linearly with the shear rate. Three types of
surface roughness were considered: thermal, random and periodic.

The thermal surface roughness due to finite spring stiffness of the
wall atoms significantly modifies the slip behavior. The large
penetration of the wall atoms into the fluid phase observed for soft
walls causes weak rate dependence of the slip length below its
values for atomically smooth rigid walls. Increasing the wall
stiffness produces effectively smoother surfaces and leads to the
linear rate dependence of the slip length. Periodically and randomly
corrugated rigid surfaces, with the amplitude below the molecular
diameter, strongly reduce the slip length and its shear rate
dependence. These findings open perspectives for modeling complex
systems with combined effects of surface roughness, wettability and
rate dependency.

\section*{Acknowledgments}
Financial support from the Michigan State University Intramural
Research Grants Program is gratefully acknowledged. Computational
work in support of this research was performed at Michigan State
University's High Performance Computing Facility.

\bibliographystyle{prsty}

\end{document}